\documentclass[a4paper,11pt]{article}
\usepackage{jheppub} 

\usepackage[T1]{fontenc} 

\newcommand{\ra}{\rightarrow}

\title{\boldmath Duality and Triality Families of Analytic Black Hole Solutions}

\author{Zhaojie Xu}
\affiliation[a]{Department of Physics, Shanghai University, \\Shanghai 200444, P.R. China}

\emailAdd{zeezj@shu.edu.cn}

\abstract{Recent progress in holographic realization of strange metal transport has given us new insights into getting new black hole solutions. In this paper, we consider the $S$-completion and $ST$-completion of the Gubser-Rocha model. The dyonic black holes of the S-type Gubser-Rocha model come in families as $\mathbb{Z}_4$ quartets. For the $ST$-completion, solutions form sextets under $\mathbb{Z}_6$ symmetry. The charge vectors of the solutions in each $\mathbb{Z}_6$-family form a hexagon, realizing six-fold way in gravitational systems.}

\makeatletter
\gdef\@fpheader{Typeset written in JHEP style}
\makeatother

\begin{document}
\maketitle
\flushbottom

\section{Introduction}
Black holes have been a rich source of ideas for physicists lasting more than a century. The ongoing active research includes studying primordial black holes in the early universe, counting black hole microstates, studying analogies between black holes and quantum systems etc. A rich variety of black holes has been explored in the context of Einstein-Maxwell-Dilaton (EMD) theories, some of which are closely related to gauged supergravities and string/M theory. A small subset \cite{Herzog:2007ij,Hartnoll:2007ip,Goldstein:2010aw,Fujita:2012fp,Chow:2013gba} of these black holes exhibit interesting symmetry properties in their Maxwell and dilaton/axion-dilaton sectors. 

Symmetry by itself is one of the guiding principles of physics, especially in particle phenomenology. And it's not surprising that it also plays an important role in black hole physics. Practically speaking, the more symmetrical the black hole solution ansatz is, the easier it is to solve. Due to the presence of the black hole, part of the symmetries of the theory is inevitably broken while the symmetries of the matter content in many cases are well preserved. What kind of remaining symmetries exist is worth studying.

Recently it was shown in \cite{Ge:2023yom} that a dyonic AdS black hole solution can be found by adding additional terms and imposing S-duality symmetry to the Gubser-Rocha model \cite{Gubser:2009qt}. It's well known in the context of gauged Type IIB supergravity that the $\text{SL}(2;\mathbb{R})$ symmetry is spontaneously broken and at special values of the axiodilaton, certain Abelian discrete symmetry remains. Though the original Gubser-Rocha model is much easier to construct from the Type IIA/M theory side, we can still use this fact to interpret the result in \cite{Ge:2023yom} as the self-duality of $\tau=\mathrm{i}$ gets modified to pairs of $\tau$ transform to each other under modular-S transformation while accompanied by non-trivial S-transformation on the 1-form in the bulk. In this case, the $\mathbb{Z}_4$ symmetry is also preserved. Pushing this idea further, it's tentative to consider another symmetry $\mathbb{Z}_6$, which could also be preserved after spontaneous breaking of $\text{SL}(2;\mathbb{R})$. In this paper, we will show that it's indeed realizable in a variant of Gubser-Rocha model. 

This paper is organized as follows. In section 2 we consider two possible generalizations of the Gubser Rocha model. For each generalization black hole solutions come in families as certain representation of finite Abelian group. In section 3 conclusions and discussions would be made. 
\section{Gubser-Rocha model revitalized}
Recent study in \cite{Ahn:2023ciq} has shown that the Gubser-Rocha model is incapable of showing all the transport anomalies of strange metal, which can be viewed as a numerical check of a more general statement \cite{Amoretti:2016cad} on the scaling properties of transport coefficients of EMD systems. One possible way to bypass this "no-go theorem" is to consider adding topological terms that were not present in \cite{Ahn:2023ciq, Amoretti:2016cad} and \cite{Gouteraux:2014hca} to the Lagrangian and figure out the scaling relations of the transport coefficients with temperature. Triggered by this, an extension of Gubser-Rocha model was proposed in \cite{Ge:2023yom}. In the new model, the Lagrangian is invariant under S-duality symmetry and the transport coefficients can be solved exactly, therefore goes beyond the $\bf{perturbative}$-$\bf{only}$ analysis on magnetotransport in \cite{Amoretti:2016cad}. The discrepancy between EMD systems and strange metal phenomenology could be partially moderated. Though it's still not clear whether Gubser-Rocha type model could describe the IR physics of strange metal, the result in \cite{Ge:2023yom} is quite intriguing for exploring the space of possible black hole solutions. In the following, we will consider two possible extensions of Gubser-Rocha model. For each extension, one finds black hole solutions that form families that transform under a discrete Abelian subgroup of $\text{SL}(2;\mathbb{Z})$.  
\subsection{$\mathbb{Z}_4$ families}
The dyonic Gubser-Rocha action \cite{Ge:2023yom} that produces $\mathbb{Z}_4$-family solution reads
\begin{equation}\label{actionZ4}
S=\int \mathrm{d}^4x\bigg(	R \,-\,\frac{3}{2}\frac{\partial_\mu\tau\partial^\mu\bar{\tau}}{(\text{Im}\tau)^2} \,-\, \frac{1}{4} e^{-\phi}\,F^2 \,+\, \frac{1}{4} \chi\,F\,\tilde{F}+\, \frac{3}{L^2}\frac{\tau\bar{\tau}+1}{\text{Im}\tau}\bigg),
\end{equation}
where $\tau\equiv\tau_1+\mathrm{i}\,\tau_2\equiv \chi+ \mathrm{i}\,e^{-\phi}$ is the axio-dilaton, $F_{\mu\nu}$ is the field strength for the vector field $A_\mu$, $\tilde{F}^{\mu\nu}=\frac{1}{2\sqrt{-g}}\epsilon^{\mu\nu\rho\sigma}F_{\rho\sigma}$ is the dual field strength with the convention $\epsilon^{txyr}=1$. For planar solutions one can also add a linear axion term \cite{Davison:2013txa} to the action
\begin{equation}
	S_{la}=-\frac{1}{2}\int \mathrm{d}^4 x \sum_{I=1}^{2}(\partial \psi_{I})^2.
\end{equation}
The equations of motion of the action (\ref{actionZ4}) are shown to be invariant under
\begin{align}\label{dtS}
	\begin{split}
		\tau&\rightarrow\tau'=-\frac{1}{\tau},\\
		F_{\mu\nu}&\ra F'_{\mu\nu}=\tau_1 F_{\mu\nu}+\tau_2 \tilde{F}_{\mu\nu},\\
		\tilde{F}_{\mu\nu}&\ra \tilde{F}'_{\mu\nu}=\tau_1 \tilde{F}_{\mu\nu}-\tau_2 F_{\mu\nu},
	\end{split}
\end{align}
and the more explicit form of the action (\ref{actionZ4}) contains the original Gubser-Rocha action
\begin{align}
	\begin{split}
	S=\int \mathrm{d}^4 x\bigg(	& R \,-\,\frac{3}{2}(\partial\phi)^2 \,-\,\frac{3}{2}e^{2\phi}\,(\partial\chi)^2\,- \frac{1}{4} e^{-\phi}\,F^2 \,\\
		+&\, \frac{1}{4} \chi\,F\,\tilde{F}+\, \frac{1}{L^2}(6\cosh\phi+3\chi^2\,e^\phi)\bigg),
	\end{split}
\end{align}
thus we call this the $S$-completion of the Gubser-Rocha model. Notice that the $S$-completion is not unique \footnote{For example, the potential $V(\tau,\bar{\tau})=-3\alpha\frac{\tau\bar{\tau}+1}{\text{Im}\tau}-3(1-\alpha)\text{Im}\tau(1+\frac{1}{\tau\bar{\tau}}) $ for $L=1$, also satisfy the property $V(-1/\tau,-1/\bar{\tau})=V(\tau,\bar{\tau})$ and $V(\tau,\bar{\tau})|_{\tau_1=0}=-6\cosh\phi.$}, our choice of the potential is in agreement with the potential arising in $\mathcal{N}=2$ gauged SUGRA in 4d, in the hope that it may be reducible from higher dimensions. 

We are now ready to solve the eoms of this system. First we are going to consider spherically symmetric case therefore we turn off the linear axion term. Using the standard procedure of calculating eoms of EMD systems, we get a new dyonic solution given by
\begin{align}\label{solE4s}
	\begin{split}
\mathrm{d} s^2&=-f(r) \mathrm{d} t^2+\frac{1}{f(r)} \mathrm{d} r^2+g(r)d^2\Omega, \\
     f &=r^{1 / 2}(r+\rho)^{3 / 2}\left(\frac{1}{L^2}+\frac{1}{(r+\rho)^2}-\frac{Q^2+P^2}{3 \rho(r+\rho)^3}\right), \\
	 g&=r^{1 / 2}(r+\rho)^{3 / 2}, \\
	 A&=Q(\frac{1}{r0+\rho}-\frac{1}{r+\rho})\mathrm{d}t+P\cos\theta\,\mathrm{d}\phi, \\
	 \tau&=\frac{Q\, P\, \rho+\mathrm{i}\left(Q^2+P^2\right) \sqrt{r(\rho+r)}}{(Q^2+P^2)r+P^2\rho}.
	 \end{split}
\end{align}
After a simple coordinate change, one can get a hyperbolic solution. The planar solution is already given in \cite{Ge:2023yom}\footnote{In order to have a unified discussion on these solutions, here's a slight abuse of notation where we denote the chemical potential as Q.}:
\begin{align}\label{solE4p}
	\begin{split}
		\mathrm{d} s^2&=-f(r) \mathrm{d} t^2+\frac{1}{f(r)} \mathrm{d} r^2+g(r)\,(\mathrm{d}x^2+\mathrm{d}y^2), \\
		f &=r^{1 / 2}(r+\rho)^{3 / 2}\left(\frac{1}{L^2}-\frac{k^2}{(r+\rho)^2}-\frac{(Q^2+P^2)(r_0+\rho)^2}{3 \rho(r+\rho)^3}\right), \\
		g&=r^{1 / 2}(r+\rho)^{3 / 2}, \\
		A&=Q(1-\frac{r_0+\rho}{r+\rho})\mathrm{d}t+P(r_0+\rho)\,x\mathrm{d}y \\
		\tau&=\frac{-Q\, P\, \rho+\mathrm{i}\left(Q^2+P^2\right) \sqrt{r(\rho+r)}}{(Q^2+P^2)r+P^2\rho},\\
		\psi_1&=k\,x,\qquad \psi_2=k\,y.
	\end{split}
\end{align}
In all the cases the axiodilaton approches the self-dual point at the boundary, i.e.
\begin{equation}
	\tau|_{r\to\infty}=\mathrm{i}.
\end{equation}
Each solution labeled by $(Q,P)$ is related to another three solutions by acting the $S$-transform (\ref{dtS}) on (\ref{solE4s}) or (\ref{solE4p}) repeatedly. To be more specific, solutions labeled by
\begin{equation}
	(Q,P)\quad(P,-Q)\quad(-Q,-P)\quad(-P,Q),
\end{equation} 
furnish a representation of $\mathbb{Z}_4$. This result simply reconfirms us dyonic black holes are not only black holes, but also dyon-like objects.

\subsection{$\mathbb{Z}_6$ families}
One can slightly modify (\ref{actionZ4}) by adding a relevant term to the action
\begin{equation}\label{actionZ6}
	S=\int d^4x\bigg(	R \,-\,\frac{3}{2}\frac{\partial_\mu\tau\partial^\mu\bar{\tau}}{(\text{Im}\tau)^2} \,-\, \frac{1}{4} e^{-\phi}\,F^2 \,+\, \frac{1}{4} \chi\,F\,\tilde{F}+\, \frac{3}{L^2}\frac{\tau\bar{\tau}+\text{Re}\tau+1}{\text{Im}\tau}\bigg),
\end{equation}
whose eoms are invariant under $ST$-transformation:
\begin{align}\label{dtST}
	\begin{split}
		\tau&\rightarrow\tau'=-\frac{1}{\tau+1},\\
		F_{\mu\nu}&\ra F'_{\mu\nu}=(\tau_1+1) F_{\mu\nu}+\tau_2 \tilde{F}_{\mu\nu},\\
		\tilde{F}_{\mu\nu}&\ra \tilde{F}'_{\mu\nu}=(\tau_1+1) \tilde{F}_{\mu\nu}-\tau_2 F_{\mu\nu}.
	\end{split}
\end{align}
Again the $ST$-completion of the original Gubser-Rocha action may not be unique , our choice seems to be the most canonical one. Though the action (\ref{actionZ6}) contains (\ref{actionZ4}), their symmetry properties are completely different and the calculations are more involved. For simplicity we first warm up with neutral solutions. For the spherical case, it turns out that the solutions are just asymptotically AdS spacetime without an event horizon, here we show one of the solutions of the $\mathbb{Z}_6$ family:
\begin{align}\label{solE6sn}
	\begin{split}
		\mathrm{d} s^2&=-f(r) \mathrm{d} t^2+\frac{1}{f(r)} \mathrm{d} r^2+g(r)d^2\Omega, \\
		f &=r^{1 / 2}(r+\rho)^{3 / 2}\left(\frac{\sqrt{3}}{2L^2}+\frac{1}{(r+\rho)^2}\right), \\
		g&=r^{1 / 2}(r+\rho)^{3 / 2}, \\
		A&=0, \\
		\tau&=-\frac{1}{2}+\frac{\sqrt{3}}{2}\mathrm{i}\sqrt{1+\frac{\rho}{r}}.
	\end{split}
\end{align}
And one of the planar solutions is given by
\begin{align}\label{solE6pn}
	\begin{split}
		\mathrm{d} s^2&=-f(r) \mathrm{d} t^2+\frac{1}{f(r)} \mathrm{d} r^2+g(r)\,(\mathrm{d}x^2+\mathrm{d}y^2), \\
		f &=\frac{1}{2}r^{1 / 2}(r+\rho)^{3 / 2}\left(\frac{\sqrt{3}}{L^2}-\frac{k^2}{(r+\rho)^2}\right), \\
		g&=r^{1 / 2}(r+\rho)^{3 / 2}, \\
		A&=0, \\
		\tau&=-\frac{1}{2}+\frac{\sqrt{3}}{2}\mathrm{i}\sqrt{1+\frac{\rho}{r}},\\
		\psi_1&=k\,x,\qquad \psi_2=k\,y.
	\end{split}
\end{align}
The other solutions can be generated by acting $ST$ on $\tau$ once and twice.\footnote{For people from the GR community who may not be familiar with group theoretic languagues, one may wonder why there are only three solutions in the family instead of 6. This is because the $\mathbb{Z}_2$ center of $\text{SL}(2;\mathbb{Z})$ doesn't act on $\tau$ and we are considering neutral solutions here. This is also well explained in \cite{Antinucci:2022vyk}.} The boundary value of $\tau$ is
\begin{equation}
	\tau|_{r\to\infty}=e^{\frac{2\pi\mathrm{i}}{3}},
\end{equation}  
which is another self-dual point of the moduli space of $\tau$. Because of the non-vanishing of the axion, the neutral solutions here are quite different from \cite{Gao:2004tu,Ren:2019lgw}, therefore the solutions here $\bf{cannot}$ be continuously deformed to the neutral solutions of standard Gubser-Rocha model.

For more general dyonic solutions, we first discuss the case in which the axiodilaton doesn't depend on the charges. The spherical solutions are given by
\begin{align}\label{solE6s}
	\begin{split}
		\mathrm{d} s^2&=-f(r) \mathrm{d} t^2+\frac{1}{f(r)} \mathrm{d} r^2+g(r)d^2\Omega, \\
		f &=r^{1 / 2}(r+\rho)^{3 / 2}\left(\frac{\sqrt{3}}{2L^2}+\frac{1}{(r+\rho)^2}-\frac{\sqrt{3}(Q_i^2+P_i^2)}{6\rho(r+\rho)^3}\right), \\
		g&=r^{1 / 2}(r+\rho)^{3 / 2}, \\
		A&=\pm Q_i(\frac{1}{r0+\rho}-\frac{1}{r+\rho})\mathrm{d} t\pm P_i\,\cos\theta \mathrm{d}\phi, \\
		\tau&=\tau_{(i)},\qquad i=0,1,2.
	\end{split}
\end{align}
And the planar solutions are given by
\begin{align}\label{solE6p}
	\begin{split}
		\mathrm{d} s^2&=-f(r) \mathrm{d} t^2+\frac{1}{f(r)} \mathrm{d} r^2+g(r)\,(\mathrm{d}x^2+\mathrm{d}y^2), \\
		f &=\frac{1}{2}r^{1 / 2}(r+\rho)^{3 / 2}\left(\frac{\sqrt{3}}{L^2}-\frac{k^2}{(r+\rho)^2}-\frac{(Q_i^2+P_i^2)(r_0+\rho)^2}{\sqrt{3}\rho(r+\rho)^3}\right), \\
		g&=r^{1 / 2}(r+\rho)^{3 / 2}, \\
		A&=\pm Q_i(1-\frac{r_0+\rho}{r+\rho}))\mathrm{d} t\mp P_i(r_0+\rho)y\,\mathrm{d}x, \\
		\tau&=\tau_{(i)},\qquad i=0,1,2,\\
		\psi_1&=k\,x,\qquad \psi_2=k\,y,
	\end{split}
\end{align}
where 
\begin{align}
	\begin{split}
 \tau_{(0)}&=-\frac{1}{2}+\frac{\sqrt{3}}{2}\mathrm{i}\sqrt{1+\frac{\rho}{r}},\\ \tau_{(1)}&=-\frac{1}{\tau_0+1}=
 -\frac{2r}{4r+3\rho}+\mathrm{i}\frac{2 \sqrt{3} \sqrt{r(r+\rho)}}{4 r+3 \rho},\\ \tau_{(2)}&=-\frac{1}{\tau_1+1}=-1+\frac{2r}{4r+3\rho}+\mathrm{i}\frac{2 \sqrt{3} \sqrt{r(r+\rho)}}{4 r+3 \rho},
 \end{split}
\end{align}
and
\begin{equation}
	(Q_0,P_0)=(Q,0),\quad (Q_1,P_1)=(\frac{1}{2}Q,\frac{\sqrt{3}}{2}Q),
\quad	(Q_2,P_2)=(-\frac{1}{2}Q,\frac{\sqrt{3}}{2}Q).
\end{equation}
The charge vectors in one family are related to each other through reflection and rotation by an angle of $2\pi/3$, thus the solutions labeled by $\{\pm( Q_i, P_i),\tau_{(i)}\}$ furnish a representation of $\mathbb{Z}_6$. 

For the case where the axiondilaton is given as a function of $Q$ and $P$, the form of the solution is still given by (\ref{solE6s}) and (\ref{solE6p}), with $\tau_{(i)}$ gets generalized to
\begin{align}
	\begin{split}
\tau_{(i)}=&	\frac{-\left(Q_i^2+P_i^2\right) r+\sqrt{3} P_i Q_i \rho-P_i^2 \rho+i \sqrt{3}\left(P_i^2+Q_i^2\right) \sqrt{r(r+\rho)}}{2\left(\left(Q_i^2+P_i^2\right) r+P_i^2 \rho\right)}\\
&i=0,1,2,
\end{split}
\end{align}
with the charge vectors related to each other by the formula
\begin{equation}
	Q_{i+1}+\mathrm{i}\,P_{i+1}=e^{\frac{\pi\mathrm{i}}{3}}(Q_{i}+\mathrm{i}\,P_{i})\quad i \,\text{mod}6,
\end{equation}
showing more explicitly that the solutions labeled by $(Q_i,P_i)$ with $i=0,\cdots,5$ can be grouped into a sextet under $\mathbb{Z}_6$. Here $(Q_{i+3},P_{i+3})=-(Q_i,P_i)$ and $(Q_i,P_i)$ corresponds to the same $\tau_{(i)}$, simply illustrating the isomorphism $\mathbb{Z}_6\simeq\mathbb{Z}_3\times\mathbb{Z}_2$.

Due to non-vanishing constant axion on the boundary, the topological term $\chi \mathrm{F}\wedge\mathrm{F}$ reduces to a $\text{U}(1)$ Chern-Simons term $a\mathrm{d}a$ on the boundary, indicating that there might be interesting applications to certain condensed matter systems. 
\section{Conclusions and discussions}
\begin{figure}[t]
	\begin{center}
		\begin{minipage}[b]{0.45\linewidth}
			\centering
			\includegraphics[height=5.9cm]{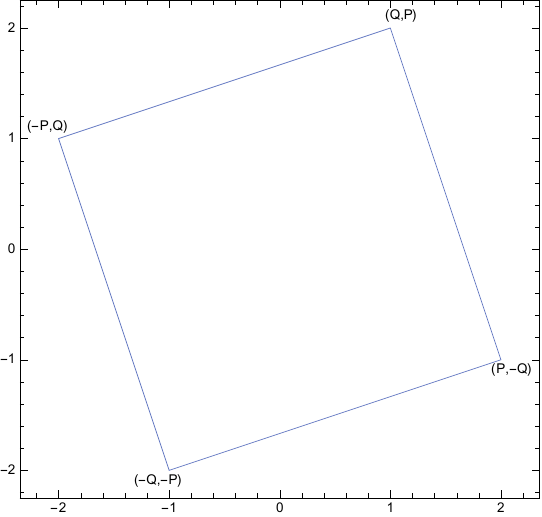}\\
			(a)
		\end{minipage} \hspace{0.2cm}
		\begin{minipage}[b]{0.45\linewidth}
			\centering
			\includegraphics[height=5.9cm]{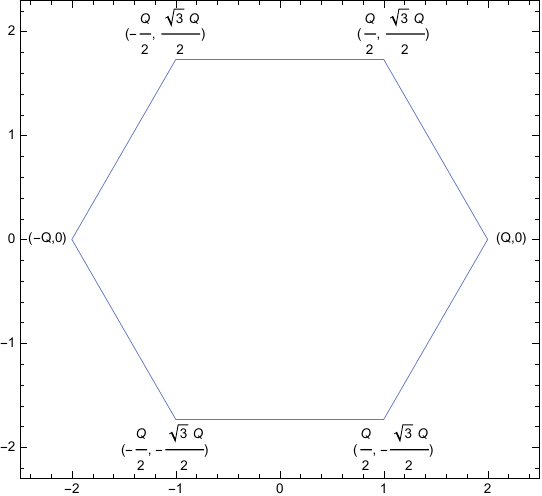}\\
			(b)
		\end{minipage}
		
	\end{center}
	\caption{charge vectors for each family of black hole solutions: (a) $S$-completed case with $(Q,P)=(2,1)$ in approapriate units, (b) $ST$-completed case with $(Q_0,P_0)=(2,0)$.}
	\label{fig1}
\end{figure}

To summarize, we have obtained families of black hole solutions in two different setup based on the $S$-completion and the $ST$-completion of the Gubser Rocha model. For the $S$-completed Gubser-Rocha model the solutions form representations of $\mathbb{Z}_4$ while for $ST$-completed Gubser-Rocha model, the solutions form representations of  $\mathbb{Z}_6$. The charge vectors of each solution family form a square and a hexagon respectively, see Fig(\ref{fig1}), realizing four-fold way and six-fold way in gravitational systems.

There are many possible extensions of this work. First one can try other potentials that are invariant under $S$ or $ST$ transformation and see if there's any interesting black hole solutions. Secondly one can consider multiple dyon case of the $\mathbb{Z}_4$ system and investigate the generalized fusion rules of dyonic black holes. Additionaly, since the concept of symmetry \cite{Gaiotto:2014kfa,Freed,Schafer-Nameki:2023jdn} has been drastically revolutionalized in the past ten years and in this paper we have restricted ourselves to the 0-form symmetry, it would be interesting to consider the generalized symmetries of the boundary QFT and its holographic realizations. Moreover, we haven't discuss the constraint on the charge lattice from the swampland weak gravity conjecture (WGC) \cite{Arkani-Hamed:2006emk,Harlow:2022ich} point of view, it would be also interesting to work this out.

\acknowledgments

We would like to thank Xian-Hui Ge, Shuta Ishigaki for helpful discussions.




\bibliographystyle{JHEP}
\bibliography{biblio.bib}
\end{document}